\documentclass[10pt,sigconf,letterpaper]{acmart}

\usepackage[english]{babel}
\usepackage{blindtext}
\usepackage{graphicx}
\usepackage{tikz}
\usepackage{color}
\usepackage{booktabs}
\usepackage{subcaption}
\usepackage[export]{adjustbox}
\usepackage{pgfplots}
\usepgfplotslibrary{dateplot}

\begin{document}
\title{Behavior of Liquidity Providers in Decentralized Exchanges}

\author{Lioba Heimbach}
\authornote{The first two authors have equal contribution.}
\email{	hlioba@ethz.ch}
\affiliation{%
  \institution{ETH Zurich}
  \city{Zurich}
  \state{Switzerland}
}

\author{Ye Wang}
\authornotemark[1]
\email{wangye@ethz.ch}
\affiliation{%
  \institution{ETH Zurich}
  \city{Zurich}
  \state{Switzerland}
}

\author{Roger Wattenhofer}
\email{wattenhofer@ethz.ch}
\affiliation{%
  \institution{ETH Zurich}
  \city{Zurich}
  \state{Switzerland}
}

\begin{abstract}
    Decentralized exchanges (DEXes) have introduced an innovative trading mechanism, where it is not necessary to match buy-orders and sell-orders to execute a trade. DEXes execute each trade individually, and the exchange rate is automatically determined by the ratio of assets reserved in the market. Therefore, apart from trading, financial players can also liquidity providers, benefiting from transaction fees from trades executed in DEXes. Although liquidity providers are essential for the functionality of DEXes, it is not clear how liquidity providers behave in such markets.
    
    In this paper, we aim to understand how liquidity providers react to market information and how they benefit from providing liquidity in DEXes. We measure the operations of liquidity providers on Uniswap and analyze how they determine their investment strategy based on market changes. We also reveal their returns and risks of investments in different trading pair categories, i.e., stable pairs, normal pairs, and exotic pairs. Further, we investigate the movement of liquidity between trading pools. To the best of our knowledge, this is the first work that systematically studies the behavior of liquidity providers in DEXes.
\end{abstract}

\maketitle

\section{Introduction}

Traditionally, 
trading is executed on centralized exchanges (CEXes), using the limit order book mechanism. With this mechanism, each seller is matched with a buyer for a trade.
In the cryptospace, when traders for instance want to exchange Ether (ETH) with Bitcoin (BTC) through a CEX, they have to transfer their ETH to the account of the centralized operator, submit their sell orders of ETH, wait for matching corresponding buy orders of ETH, and then withdraw the incoming BTC from the market operators after the execution of their orders. This is cumbersome but also risky, since the funds are temporarily with the CEX.

Recently, there is an alternative in the form of decentralized exchanges (DEXes). Currently DEXes are popular in cryptocurrency markets, but eventually fiat currencies, stocks or other commodities might be traded on DEXes as well. So understanding DEXes may be important beyond just cryptocurrency applications.

DEXes are based on smart contracts running on a blockchain. In contrast to the limit order book mechanism, traders do not need to be matched to a trading partner with opposite intentions. In a DEX, trades are completed immediately when the orders are recorded on the blockchain. 

Uniswap is the most popular 
DEX. As an example, consider trades between BTC and ETH. The Uniswap platform offers a smart contract (liquidity pool) with locked-in funds for these two cryptocurrencies. When a trader wants to exchange ETH for BTC, the trader just needs to send their ETH to the smart contract. The smart contract will then immediately send the appropriate amount of BTC back to the trader, while the ETH sent by the trader will be locked in the smart contract. The exchange rate is primarily determined by the ratio of BTC to ETH stored in the smart contract. 
The BTC and ETH in the smart contract are providing the liquidity for the trades between BTC and ETH on Uniswap. 

In a DEX pool, liquidity is provided by liquidity providers. These liquidity providers lock their cryptocurrencies into the corresponding liquidity pools. 
DEXes generally charge a percentage transaction fee for each trade executed on the platform. This transaction fee is shared by the liquidity providers in proportion to their liquidity contributions.  Therefore, with the emergence of DEXes, users have new investment opportunities in the cryptocurrency ecosystem: they can offer their assets in DEXes as liquidity and benefit from transaction fees. 

Many interesting questions emerge: How many different pools are liquidity providers invested in? What are the expected earnings from transaction fees? How do liquidity providers react to market forces? Do liquidity providers redistribute their assets? As DEXes are an emerging phenomenon that might cross over to other markets beyond cryptocurrencies, it is interesting to understand the motives of liquidity providers. 


In this paper, we quantitatively measure the behavior of liquidity providers. We collect data from Uniswap, the most popular DEX in the cryptocurrency ecosystem, to study liquidity provider contributions, returns, and strategies. First, we analyze the distribution of liquidity providers in Uniswap, examining the creation of liquidity pools, the distribution of liquidity, and the participation of liquidity providers in these pools. Although Uniswap allows users to create liquidity pools between any pair of tokens, more than 80\% of liquidity pools include ETH, and six popular tokens dominate the market. Moreover, more than 60\% liquidity is locked in the top 24 pools. Cryptocurrency holders act restrained concerning these new investments: approximately 70\% of providers reserve their liquidity in a single pool. However, these seemingly conservative liquidity providers contribute more than 50\% of the liquidity in the most popular pools. This indicates that individual providers, as opposed to professional market makers, control the liquidity of DEXes. In examining the addition and removal of liquidity day by day, we find that the market change of liquidity is relatively stable, i.e., the correlation between the number of injections and withdrawals is 0.922. 


To understand the behavior of liquidity providers, we study the risks and returns they face across three different pool Uniswap categories: stable, normal and exotic. We find that while stable and normal pairs may provide attractive investment opportunities for liquidity providers depending on the individual risk tolerance and return expectations, the exotic pools investigated are not. Those demonstrated deeply negative returns accompanied by high risk. 


Liquidity providers do not frequently move their assets across different pools, seemingly indifferent to price changes or other market indicators. However, we observe that many traders redistribute their liquidity given the opportunity to earn additional benefits of liquidity mining on top of transaction fees. This inspires to study what external factors are influencing the behavior of liquidity providers.

This paper makes the following contributions.
First, we conduct a systematic investigation on the liquidity providers in Uniswap to outline the participants of users in such emerging trading activities.
Second, we classify three categories of liquidity pools and analyze the investment returns and risks of liquidity providers in those pools. We present the variability of investment strategies for different types of liquidity pools.
Finally, we demonstrate the movement of liquidity around the entire DEX and suggests the significant influence of external factors on liquidity providers, such as liquidity mining activities.

\section{Background and Related Work}

In this section, we introduce the background of DEXes. First, we present the technology basis of the Ethereum blockchain, smart contracts, and the mechanism of DEXes. We also review two types of previous work: quantifying user behavior on DEXes and analyzing returns of liquidity providers.

\subsection{Ethereum Blockchain and ERC20 Standard}

Ethereum is a public blockchain platform, which supports Turing complete smart contracts. Compared to earlier blockchain systems, such as Bitcoin, Ethereum provides a decentralized virtual machine, the Ethereum Virtual Machine (EVM), to execute smart contract code. A smart contract is a set of programs written in high-level languages, e.g., Solidity. These programs will be compiled into executable byte-code. After the creation of the smart contract, the executable byte-code will be stored in an independent database of the blockchain. Any Ethereum users can invoke functions defined in a smart contract.

Ethereum supports three kinds of transactions: a \textit{simple transaction}, where the recipient is another address (account) to transfer the native currency, ETH; a \textit{smart contract creation transaction}, without recipient is the \textit{null} to create a new smart contract; a \textit{smart contract execution transaction}, where the recipient is a smart contract address to execute a specific function of that contract.

When a transaction is included in a block by a miner, the operation corresponding to the message takes effect. The miner who creates the block modifies the state of corresponding accounts based on the messages. Each step of the miner's operation consumes a certain amount of gas, and the amount of gas consumed in each block is capped. Users need to specify a gas price for the operation execution when sending transactions. The fee paid by the initiator of a transaction to miners is determined by the amount of gas consumed and the gas price (gas fee = gas price $\times$ gas consumption). The miner will include a receipt of the executed transactions in Ethereum blocks as well, including the information on whether the transaction has been executed successfully, the gas fee, the identity of the transaction and the block, and other information generated during the execution.

Based on the support of the smart contract, users can create cryptocurrency other than ETH on Ethereum. These smart contracts have to follow some standards; the most widely used standard is the ERC20 standard, which requires an \texttt{approve} function and a \texttt{transferFrom} function. When an address (account) $addr_a$ calls the function \texttt{approve($addr_b, v$)}, then the address $addr_b$ can transfer at most $v$ tokens in total from $addr_a$ to other accounts. After this approval, $addr_b$ can transfer $v'$ tokens from $addr_a$ to another account $addr_c$ by calling the function \texttt{transferFrom($addr_a$, $addr_c$, $v'$)}, where $\sum v'\leq v$.

\subsection{DEXes}

DEXes are smart contracts on Ethereum. Users send messages to a DEX address to invoke functions for performing market operations. DEXes support these operations: create the trading (liquidity) pool between a pair of tokens, adding/removing liquidity, and exchanging tokens. We take Uniswap as an example to present these operations. Uniswap is often also called an automated market maker DEX or a constant function market maker DEX. There are currently three Uniswap versions, V1, V2, and V3. Uniswap V2 is the predominated market at the time of this writing. So we study Uniswap V2 as example in this paper as it provides us with the biggest data set. Furthermore, other DEXes, e.g. Sushiswap, have similar mechanisms as Uniswap V2.

\subsubsection{Creating Liquidity Pools}

In Uniswap, exchanges between two tokens are conducted through a liquidity pool, i.e., a smart contract that keeps the pair of tokens. There are two participants involved in the market: the liquidity provider and the trader. Providers reserve their tokens in the liquidity pool, while traders exchange their tokens with the liquidity pool. Because providers contribute to the market liquidity, they benefit from the transaction fees incurred with transactions in DEXes.

Assume we have two tokens $A$ and $B$, and we want to create a liquidity pool between $A$ and $B$ on Uniswap. We first send a smart contract execution transaction with the ERC20 smart contract address of $A$ and $B$ to the Uniswap to claim the creation of the liquidity pool $A\rightleftharpoons B$. The smart contract will then check whether the pool between $A$ and $B$ exists according to the addresses of two tokens. If not, Uniswap smart contract will create a new liquidity pool $A\rightleftharpoons B$, i.e., a new smart contract reserving these two tokens.

\subsubsection{Adding/Removing Liquidity}

After creating the liquidity pool between $A$ and $B$, liquidity providers can add a token pair to the liquidity pool. Liquidity providers need to approve the Uniswap address to transfer their $A$ token and $B$ token from their address to the liquidity pool address. When a Uniswap contract receives a liquidity providers call to add liquidity, it will invoke the \texttt{transferFrom} function in ERC20 contracts to transfer tokens from the provider's address to the liquidity pool address.

If there are no tokens reserved in the liquidity pool, users can supply any amount of $A$ and $B$ to the liquidity pool, and the pool will return \textit{liquidity} tokens as proof of the deposit. If the amounts of $A$ and $B$ provided by the providers are $a$ and $b$, respectively, then the provider will get $\lambda = \sqrt{a\times b}$ \textit{liquidity} tokens. Meanwhile, the total supply of \textit{liquidity} tokens of $A\rightleftharpoons B$ pool is $\Lambda=\sqrt{a\times b}$.

With $a$ of token $A$ and $b$ of token $B$ already in the liquidity pool, a provider can reserve $\delta_a$ of its asset $A$ and $\delta_b$ of its asset $B$ in the liquidity pool simultaneously, where $\frac{\delta_a}{\delta_b}=\frac{a}{b}$. Then they will earn $\delta_{\lambda} = l\cdot\frac{\delta_a}{a}$ \textit{liquidity} tokens for the $A\rightleftharpoons B$ and the total supply of \textit{liquidity} tokens becomes $\Lambda = \lambda+\delta_\lambda$.

Providers can also remove their tokens from the liquidity pool. The amount of tokens providers can redeem is related to the amount of \textit{liquidity} tokens they own. Assume a provider has $\delta_\lambda$ \textit{liquidity} tokens of the liquidity pool $A\rightleftharpoons B$ and the total supply of \textit{liquidity} tokens is $\lambda$. The provider can withdraw $\delta_a$ of $A$ and $\delta_b$ of $B$ from the $A\rightleftharpoons B$ pool with $\delta_\lambda'\leq\delta_\lambda$ \textit{liquidity} token, where $\frac{\delta_a}{a}=\frac{\delta_b}{b}=\frac{\delta_{\lambda}'}{\lambda}$. The $\delta_{\lambda}'$ of \textit{liquidity} tokens will be burned (destroyed) after they redeem the money and the total supply of \textit{liquidity} tokens becomes $\Lambda=\lambda-\delta_\lambda'$.

\subsubsection{Exchanging Assets}

In Uniswap, tokens are not exchanged between two traders but between a trader and the liquidity pool. 
The exchange of assets is realized in two steps.
First, the traders sends their tokens to the liquidity pool.
Second, the liquidity pool computes the exchange rate and returns the targeted token to the traders.

Assume traders wants to exchange $\delta_a$ of $A$ for $B$ token and the liquidity of $A$ and $B$ are $a$ and $b$. 
They first need to let the Uniswap address get the approval for transferring his $A$ token to other accounts. 
After receiving the exchange order from the traders, the Uniswap contract transfers $\delta_a$ of $A$ to the liquidity pool address and returns $\delta_b$ of $B$ back to the traders.

The following equation always holds during the exchange: $a\cdot b = (a+\delta_a\cdot r_1)\cdot(b-\frac{\delta_b}{r_2})$, where $r_1$ and $r_2$ denote the transaction fee ratio in asset $A$ and $B$ respectively. 
In Uniswap, $r_1  = 0.997$ and $r_2=1$, which indicates that the transaction fee is equal to $3$\textperthousand$\cdot \delta_a$. The remaining liquidity in the pool equals to $(a+\delta_a, b-\frac{r_1\cdot r_2\cdot b\cdot\delta_a}{a+r_1\cdot\delta_a})$ and the amount of \textit{liquidity} tokens does not change.

\subsection{Liquidity in DEXes}

This paper aims to study liquidity providers in DEXes quantitatively. In this section, we review previous work in two aspects: quantitative studies of market behavior in DEXes, and studies on liquidity providers in DEXes.

Since all transaction information is broadcast through the blockchain network and all transactions in DEXes are public to all market participants, researchers are able to study the market behavior of traders in cryptocurrency ecosystems. Chen et al.~\cite{chen2020traveling} provide a basic view of the ERC20 tokens on Ethereum. They visualize how cryptocurrencies are created, held, and transferred by traders. Other works explore more detailed behavior of traders in the markets. Daian et al.~\cite{daian2020flash} measures front-running trades on DEXes. Because of the transparency and latency of DEXes transactions, traders are able to observe profitable transactions before they are executed and place their own orders with higher fees to front-run the target victim. Such behaviors result in a high miner-extractable value, which brings systemic consensus-layer vulnerabilities. Torres et al.~\cite{torres2021frontrunner} and Qin et al.~\cite{qin2021quantifying} quantify the revenue of traders who conduct a combination of front-running and back-running transactions, i.e., sandwich attacks~\cite{zhou2020high}, demonstrating the normal transactions are vulnerable to arbitrageurs. Wang et al.~\cite{wang2021cyclic} study another arbitrage behavior, i.e., cyclic arbitrages (triangular arbitrages) in Uniswap, which profits traders with price differences across different trading pools. They claim that implementing transactions with private smart contracts is more resilient to front-running attacks than directly calling public functions of DEXes smart contracts. Although previous studies have provided an in-depth understanding of traders in the cryptocurrency ecosystems, they mainly focus on the behavior of normal traders. The novel trading option of providing liquidity on DEXes has not been well studied. Therefore, this paper aims to fill the research gap and inspires more work in this direction.

A separate line of work has analyzed the returns of liquidity providers theoretically with microstructure models. Evans~\cite{evans2020liquidity} studies the returns of liquidity providers when there is no transaction fee charged and claims that it is better to invest in constant-mix portfolios than providing liquidity. Moreover, Evans et al.~\cite{evans2021optimal} develop a framework for determining the optimal transaction fees for AMM DEXes and show that providing liquidity to DEXes is preferable to all alternative trading strategies as fees approach zero. Aoyagi ~\cite{aoyagi2021liquidity} conducts a game-theoretical analysis between liquidity providers and informed traders to estimate the returns of liquidity providers in DEXes. These studies have ideal assumptions on information delivery and price fluctuations of assets, which do not coincide with the real scenarios. Moreover, they have not considered how liquidity providers choose between different trading pools.
To the best of our knowledge, none of the previous studies have empirically analyzed the trading behavior of liquidity providers in real DEXes.
To better understand the new trading option in DEXes, we measure the real behavior of liquidity providers in Uniswap and study how they react to market indicators, which will inspire better theoretical models for analyzing liquidity providers in DEXes.

\section{Data Description}

To measure the behavior of liquidity providers in Uniswap and understand how their trading strategies are influenced by market factors, we collect all transactions recorded on Ethereum from block 10000835 (where Uniswap V2 has been deployed, on May 4, 2020) to block 11709847 (on January 23, 2021)\footnote{Block data is available at https://www.ethereum.org/}. We develop and launch a modified version of go-ethereum client, which exports all transactions executed on Uniswap.

The addresses of liquidity pool smart contacts are stored in the \textit{UniswapV2Factory} contract. We query this contract to get the contract addresses and token pairs of 29,235 available trading pools until January 23, 2021. Then, we find 21,830,282 transactions interacting with these liquidity pools as Uniswap trades and filter those containing \textit{Mint} and \textit{Burn} events, representing adding liquidity, and removing liquidity, respectively.

In each \textit{Mint} event, \textit{liquidity} tokens of the liquidity pool will be generated by the liquidity pool smart contract and then transferred from the \textit{0} address to the address of liquidity providers. Similarly, in each \textit{Burn} event, liquidity providers transfer their \textit{liquidity} tokens to the \textit{0} address. Moreover, some traders may exchange \textit{liquidity} tokens in other transactions, while liquidity pools will generate \textit{Transfer} events to record such \textit{liquidity} token movements. With the information of \textit{Mint}, \textit{Burn}, and \textit{Transfer} events, we compute the balance of \textit{liquidity} tokens of liquidity providers in each pool.

To further evaluate the behavior of liquidity providers and its dependence on pool characteristics on Uniswap, we require the price of each cryptocurrency in a common currency -- USD in our case. As many cryptocurrencies do not have a market price, we first compute the price of each cryptocurrency studied in ETH. This data is obtained from the common pool with ETH of the cryptocurrency in question. Such a pool exists for most cryptocurrencies. The ETH price then allows us to calculate the value of each token in USD historical data for the ETH price in USD from coinbase\footnote{https://www.coinbase.com/}.

\section{Liquidity Dynamics in DEXes}

\begin{figure}
\centering
\setcaptionwidth{\linewidth}
    \input{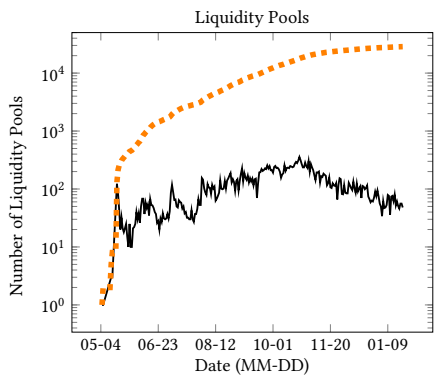}
    \caption{Number of liquidity pools on Uniswap from 4th May 2020 to 23rd January 2021. The orange dotted line represents the number of all liquidity pools on Uniswap, while the black line shows the number of daily emerging pools.}
    \label{fig:pool_num}
\end{figure}

\begin{figure}
\centering
\includegraphics[width=\linewidth]{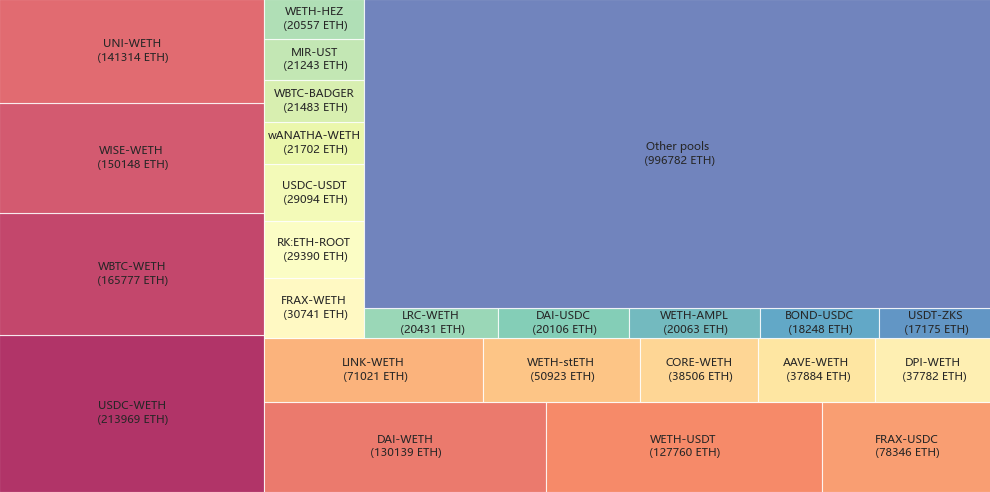}
    \caption{Distribution of liquidity in Uniswap on January 23, 2021.}
    \label{fig:num}
\end{figure}

In this section, we provide an overview of liquidity pools on Uniswap. Unlike traditional centralized exchanges where only the cryptocurrencies permitted by operators can be traded on the platforms, DEXes allow any Ethereum users to create trading pools between any pair of tokens. As shown in \autoref{fig:pool_num}, the number of liquidity pools increases quickly with the emergence of Uniswap and continues to rise at a rate of over 100 per day until November 2020. After that, the daily growth of liquidity pools has gradually slowed down and keeps at a rate of 50 pools per day by the end of January 2021. In total, we find 29,235 available liquidity pools, which involve 25,231 cryptocurrencies, while 22,828 tokens only have a single trading pool. The most popular token is ETH, 24,011 tokens share a liquidity pool with ETH. Except for ETH, the most popular cryptocurrencies are USDT (1321), USDC (627), DAI (542), UNI (270), and WBTC (148). These tokens are either stable coins whose value is anchored at 1 USD or tokens with high value. Until 23rd January 2021, Uniswap has reserved more than 2.5 billion ETH and more than 60\% of liquidity is located in the top 24 popular pools (\autoref{fig:num}).

\begin{figure}
\centering
\setcaptionwidth{\linewidth}
    \input{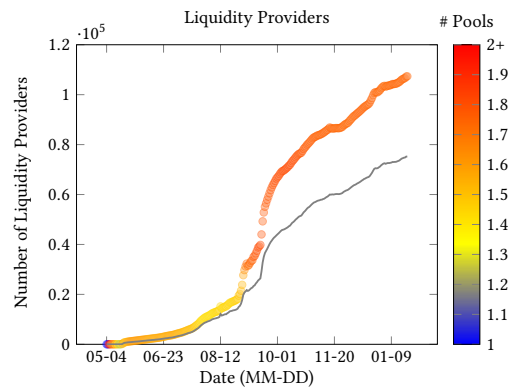}
    \caption{The number of liquidity providers over time. The gray line demonstrates the number of providers that only participate in one pool. The colorful dotted line is the number of active liquidity providers in Uniswap, and the color indicates the average number of pools that liquidity providers participate.}
    \label{fig:lp}
\end{figure}

Apart from the distribution of liquidity pools, we also measure the participation of liquidity providers in Uniswap.
Although some traders may use several addresses to provide liquidity on Uniswap, more than 82\% of Ethereum users only control a single account~\cite{victor2020address}. Therefore, in this paper, we consider each address as a single liquidity provider.
In total, we find 183,823 addresses add liquidity to liquidity pools on Uniswap, while 107,352 of them keep reserving their tokens in Uniswap by 23rd January 2021 (\autoref{fig:lp}). Initially, most liquidity providers reserve their money only in one liquidity pool, and the average number of pools they participate in is less than 1.4 until September 2020. Later on, during October and November, liquidity providers are interested in more pools, as we see the average number of pools providers are involved in increases to 1.8.

\begin{figure}
\centering
\includegraphics[width=0.9\linewidth]{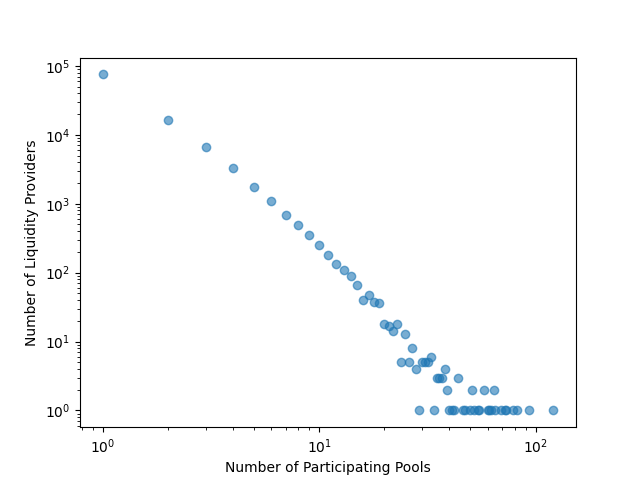}
    \caption{Distribution of the number of liquidity pools that each liquidity provider participates in. About 10,000 only participate in a single pool (top left), and some providers participate in many pools (bottom right).}
    \label{fig:powlaw}
\end{figure}

As shown in \autoref{fig:powlaw}, the distribution of the number of liquidity pools that each liquidity provider participates in follows the power law. The largest providers participate in 120 pools and more than 96.5\% of the accounts reserve their money in no more than 5 pools.

\begin{figure}
\centering
\includegraphics[width=0.9\linewidth]{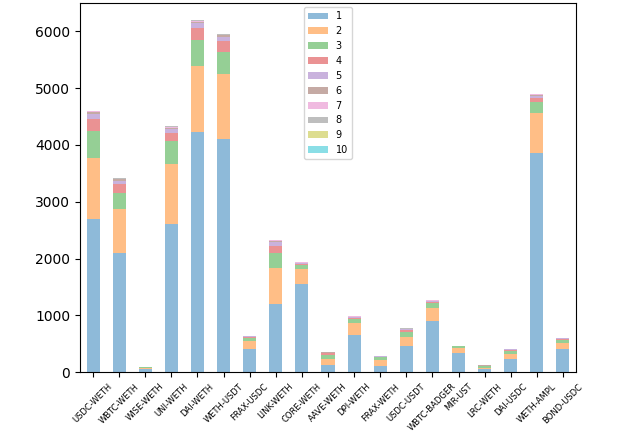}
    \caption{
    Number of liquidity providers per pool. The DAI-WETH pool has more than 6,000 individual liquidity providers. More than 4,000 of them only contributed to this one (1) pool. The other colors represent providers which are providing liquidity to more than one (2,3,...) pool.
    }
    \label{fig:pools}
\end{figure}

\begin{figure}
\centering
\includegraphics[width=0.9\linewidth]{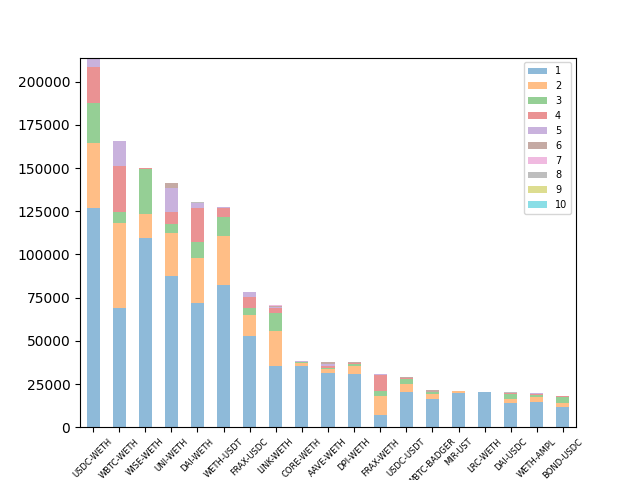}
    \caption{Total liquidity per pool. The biggest pool is USDC-WETH with more than 200,000 ETH liquidity. About 125,000 ETH is funded by liquidity providers which only contributed to this one (1) pool. The other colors represent providers which are providing liquidity to more than one (2,3,...) pool.
    }
    \label{fig:pools2}
\end{figure}

In general, becoming liquidity provider is not yet a popular investment strategy in the cryptocurrency ecosystems. Most liquidity providers only participate in a single pool. Even for those popular liquidity pools on Uniswap, more than 70\% of providers reserve their cryptocurrency assets only in one of them. Only few providers spread their assets in more than 10 pools (\autoref{fig:pools}). Liquidity providers reserve their cryptocurrency assets only in one pool have taken a serious role in the DEXes ecosystem, as they provide more than half liquidity in these popular pools (\autoref{fig:pools2}). This fact suggests that DEXes are not controlled by oligopoly and professional market makers. Ordinary users contribute the most to the healthy operation of the decentralized market mechanism.

\begin{figure}
    \centering
    \includegraphics[width=0.9\linewidth]{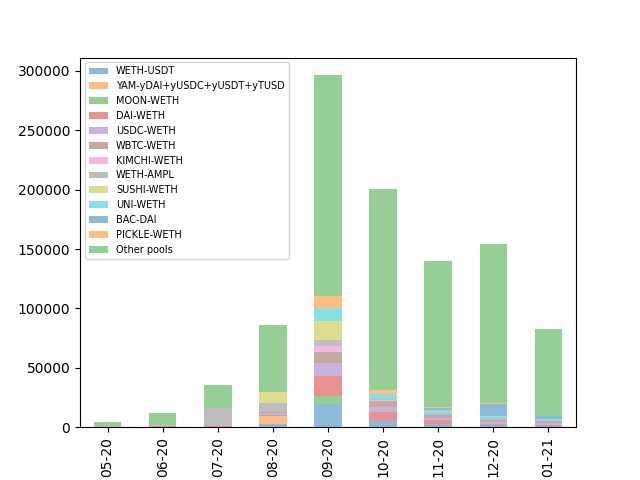}
    \caption{Number of mint events per month executed in different liquidity pools in Uniswap. In September 2020 there were more then 300,000 mint events in total.}
    \label{fig:add}
\end{figure}

\begin{figure}
    \centering
    \includegraphics[width=0.9\linewidth]{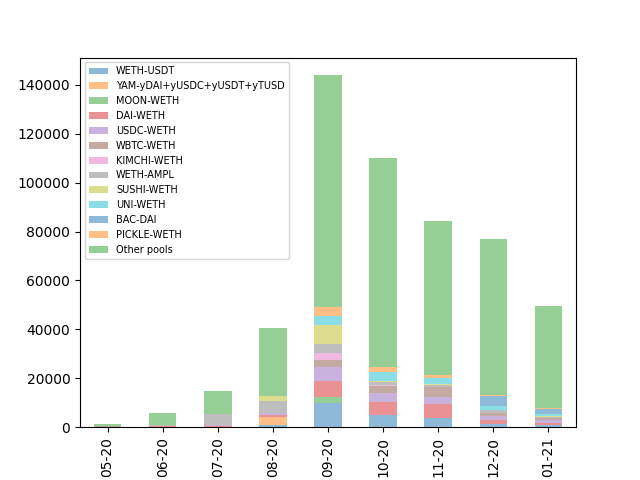}
    \caption{Number of burn events per month executed in different liquidity pools in Uniswap. In September 2020 there were more then 140,000 burn events.}
    \label{fig:remove}
\end{figure}

During the nine analyzed months, liquidity providers have gradually added and removed liquidity from different pools, while they reserve their tokens 1,011,524 times on Uniswap (\autoref{fig:add}) and withdraw them for 527,429 times (\autoref{fig:remove}). We find that the day by day correlation between the number of liquidity injections and liquidity withdrawals is very high (0.992), indicating that the market size is growing steadily. Moreover, in terms of days, the market behaviors of liquidity providers are relatively consistent.
Almost half of liquidity operations from August to September take place in the most popular pools. However, these pools become less active from October onwards. On the one hand, we may consider the liquidity in these popular pools to become stable after five months of the development. On the other hand, this fact also indicates a rapid change of interests in the liquidity pools of investors in the cryptocurrency ecosystems.

We have specified 12 pools with the most liquidity injections and withdrawals in \autoref{fig:add} and \autoref{fig:remove}. As most liquidity operations occur in these pools, the patterns of liquidity changes in these pools can represent the trading behavior of liquidity providers in DEXes well. For instance, we can observe a clear difference between the three types of liquidity pools: normal pools, stable pools, and exotic pools.

In the normal liquidity pools, both cryptocurrencies traded in the pools have a certain value, such as $USDC\rightleftharpoons WETH$, $WBTC\rightleftharpoons WETH$, and $DAI\rightleftharpoons WETH$. These cryptocurrencies are recognized in the cryptocurrency ecosystems. To support trades between these cryptocurrencies, the liquidity in the pools is relatively high (\autoref{fig:num}). Although the price of normal tokens may fluctuate, the price trend is relatively stable and in line with the development of the cryptocurrency ecosystem.

In stable pools, both tokens traded in the pool are stable coins, such as $USDC\rightleftharpoons USDT$ and $DAI\rightleftharpoons USDC$. The price fluctuations of stable coins are negligible. Thus, the market environment has little impact on stable pool liquidity providers. Since the price of the stable coins remains largely constant, liquidity providers earn profits by charging transaction fees.

The remaining pools are referred to as exotic pools. In such pools the price of one trading token 
is extremely volatile. The price of these tokens changed by more than a hundred times during our measuring period. As we show in \autoref{fig:num}, although many liquidity operations have been applied on these pools, they are not part of the high liquidity value by 23rd January 2021, as little liquidity remains in the pool -- for example, YAM, MOON, and KIMCHI.
For these exotic cryptocurrencies, liquidity providers take more risks because of their dramatic price fluctuations. When the price of the exotic token changes drastically, the other coin they have reserved in the pool may be emptied instantly by other traders.

Given the significant differences of these three kinds of pools, we infer that the distribution of liquidity providers, their trading strategies, and the investment return across these three categories differ. Therefore, we measure the activities of liquidity providers in these pools separately and analyze how they react market changes.

\section{Returns and Risks of Providing Liquidity}

Evans et al.~\cite{evans2020liquidity} suggest that the returns of providing liquidity are lower than investing in the constant-mix portfolios if the transaction fee is zero. However, in reality, liquidity providers can benefit from their contribution to the liquidity pools through strictly positive transaction fees, which results in different returns and risks compared to the previous theoretical analysis. To better understand liquidity providers' motivation in DEXes, we compare the return received by liquidity providers to holding the respective assets according to the given token ratio during the initial liquidity injection.

The return of a liquidity provider between time $t_1$ and $t_2$ in percent is given as 
$$ \text{return}_{t_1 \rightarrow t_2}= 100\cdot\frac{\frac{\text{invest}_{t_2}}{\text{hold}_{t_2}}-\frac{\text{invest}_{t_1}}{\text{hold}_{t_1}}}{\frac{\text{invest}_{t_1}}{\text{hold}_{t_1}}},$$
where $\text{invest}_{t}$, which is the current value in USD of the liquidity placed in the pool at time $t$ and $\text{hold}_{t}$ is the value in USD of the constant-mix portfolio at time $t$.

This return is positively influenced by the fees collected from trades performed in the pool and negatively impacted by the \textit{impermanent loss}, otherwise referred to as divergence loss.  Consider a liquidity pool $A\rightleftharpoons B$ between token $A$ and $B$, where the amount of $A$ in $A\rightleftharpoons B$ at time $t$ is denoted as $a_t$ and the amount of $B$ in $A\rightleftharpoons B$ at time $t$ is denoted as $b_t$. The fees collected between time $t_1$ and $t_2$ as a percentage of the liquidity are given as,
$$\text{fees}_{t_1 \rightarrow t_2}=100\cdot\left(1-\frac{\sqrt{k_{t_1}}}{\sqrt{k_{t_2}}}\right),$$
where $k_{t} =a_t\cdot b_t$ \cite{adams2020uniswap}. The impermanent loss, on the other hand, describes the risk for liquidity providers of seeing the value of their reserved tokens decrease in comparison to holding the assets. This occurs with any price change in the pool. More precisely, the impermanent loss between $t_1$ and $t_2$ is given as,
$$\text{impermanent loss}_{t_1 \rightarrow t_2}=100\cdot\left(\frac{2\cdot \sqrt{\frac{p_{t_2}}{p_{t_1}}}}{1+\frac{p_{t_2}}{p_{t_1}}}-1\right),$$
where $$p_{t}=\frac{b_t}{a_t},$$ is the ratio between tokens in the pool at time $t$ \cite{UniswapWeb}. Due to the impermanent loss, reserving of tokens runs the risk of under-performing a buy and hold strategy of a constant-mix portfolio. Note that we have to include impermanent loss in the 

\begin{figure}
  \begin{subfigure}{\linewidth}
    \includegraphics[scale=0.53,right]{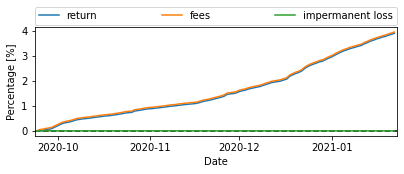}
    \caption{$USDC\rightleftharpoons USDT$} \label{fig:USDC-USDT}
  \end{subfigure}%
    
  \begin{subfigure}{\linewidth}
    \includegraphics[scale=0.53,right]{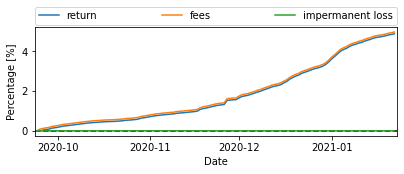}
    \caption{$DAI\rightleftharpoons USDT$} \label{fig:DAI-USDT}
  \end{subfigure}%
  
  \begin{subfigure}{\linewidth}
    \includegraphics[scale=0.53,right]{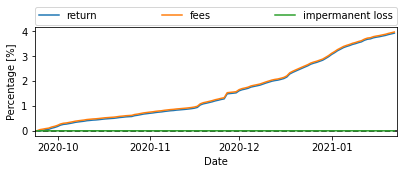}
    \caption{$DAI\rightleftharpoons USDC$} \label{fig:DAI-USDC}
  \end{subfigure}

\caption{Evolution of cumulative returns, fees and impermanent loss over four months for stable pairs. Stable pairs do not suffer from impermanent loss.} \label{fig:evstable}
\end{figure}

The returns and the impermanent loss may vary greatly across the different liquidity pool categories, namely normal pools, stable pools, and exotic pools. Thus, to provide a comprehensive understanding of returns and risks of providing liquidity in Uniswap, we analyze them separately. We look at nine pools in detail - three of each kind (stable pairs: USDC$\rightleftharpoons$ USDT, DAI$\rightleftharpoons$USDT, and DAI$\rightleftharpoons$USDC; normal pairs: UNI$\rightleftharpoons$WETH, LINK$\rightleftharpoons$WETH, DPI$\rightleftharpoons$WETH; exotic pairs: MOON$\rightleftharpoons$WETH, KIMCHI$\rightleftharpoons$SUSHI, KIMCHI$\rightleftharpoons$ WETH)\footnote{In picking pools, we excluded liquidity mining pools (i.e., WETH$\rightleftharpoons$USDT, WETH$\rightleftharpoons$USDC, WETH$\rightleftharpoons$DAI and WETH$\rightleftharpoons$WBTC), as the influence of the liquidity mining program clearly presents itself in the data.}. We choose pools through a combination of size and variety. When analyzing the data, we observe the same four-month period between the end of September 2020 and the end of January 2021, i.e., the time during which all nine sample pools were active. Further, when looking at daily returns, we consider the average daily return as opposed to the closing daily return to mitigate the effects of short-term in-balances in pool reserves. Such imbalances occasionally occur after trades that are large in comparison to the pool reserves. Due to the influence of the impermanent loss on the returns, the inaccurate price ratio causes starkly negative returns. However, these temporary imbalances recover quickly as other trades see an arbitrage opportunity due to the inaccurate price ratio. 

\begin{figure}
  \begin{subfigure}{\linewidth}
    \includegraphics[scale=0.53,right]{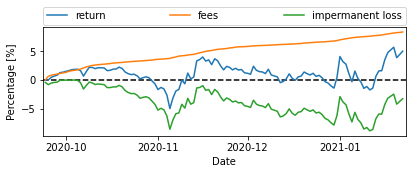}
    \caption{$UNI\rightleftharpoons WETH$} \label{fig:UNI-WETH}
  \end{subfigure}%
    
  \begin{subfigure}{\linewidth}
    \includegraphics[scale=0.53,right]{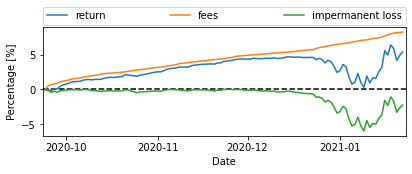}
    \caption{$LINK\rightleftharpoons WETH$} \label{fig:LINK-WETH}
  \end{subfigure}%
  
  \begin{subfigure}{\linewidth}
    \includegraphics[scale=0.53,right]{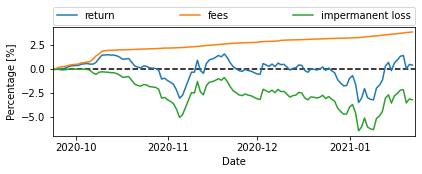}
    \caption{$DPI\rightleftharpoons WETH$} \label{fig:DPI-WETH}
  \end{subfigure}

\caption{Evolution of cumulative returns, fees and impermanent loss over four months for normal pairs. Somebody funding LINK-WETH in November, and pulling out in January would have suffered a loss, since the accumulated fees do not make up for the impermanent loss.} \label{fig:evnormal}
\end{figure}

\begin{figure}
  \begin{subfigure}{\linewidth}
    \includegraphics[scale=0.53,right]{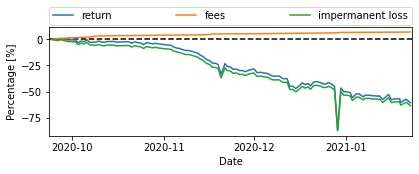}
    \caption{$MOON\rightleftharpoons WETH$} \label{fig:MOON-WETH}
  \end{subfigure}%
    
  \begin{subfigure}{\linewidth}
    \includegraphics[scale=0.53,right]{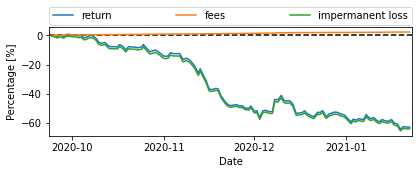}
    \caption{$KIMCHI\rightleftharpoons SUSHI$} \label{fig:KIMCHI-SUSHI}
  \end{subfigure}%
  
  \begin{subfigure}{\linewidth}
    \includegraphics[scale=0.53,right]{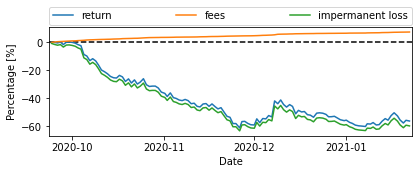}
    \caption{$KIMCHI\rightleftharpoons WETH$} \label{fig:KIMCHI-WETH}
  \end{subfigure}

\caption{Evolution of cumulative returns, fees and impermanent loss over four months for exotic pairs. The fees should be much higher considering the risk of impermanent loss.} \label{fig:evexotic}
\end{figure}

In ~\autoref{fig:evstable},~\ref{fig:evnormal} and~\ref{fig:evexotic}, we show the evolution of the return, fees and impermanent loss of stable pools, normal pools, and exotic pools over time, respectively. Each of the three stable coins (USDT, USDC, and DAI) tracks the USD. Thus, the ratio between each pair is close to one at all times, and impermanent loss plays a subordinate role in determining the returns and is therefore negligible. Rather, the fees received through trades in the pool dominate the return. This dependency is apparent from the almost overlaying curves for return and fees shown across all three pairs show in \autoref{fig:evstable}. Moreover, due to negligible price fluctuations and continuous collection of fees, liquidity providers in stable pools can expect positive returns independent of the time of the liquidity injection. Therefore, providing liquidity in stable pools appears almost risk-free with consistent and stable revenue.


The influence of the impermanent loss on the profits of liquidity providers becomes more apparent for the two other types of pools -- normal and exotic. For normal pairs (\autoref{fig:evnormal}), the cumulative return fluctuates below and above zero, influenced both by the ever-changing impermanent loss and the steadily increasing fees collected. Finally, we observe even starker domination of the impermanent loss for the return rate in \autoref{fig:evexotic}. Due to the high price volatility characteristic for exotic pairs, we observe impermanent losses of around 70\% over a four-month period, which the collected fees cannot compensate -- leading to deeply red returns. 

\begin{figure}
  \begin{subfigure}[b]{\linewidth}
    \includegraphics[scale=0.53,right]{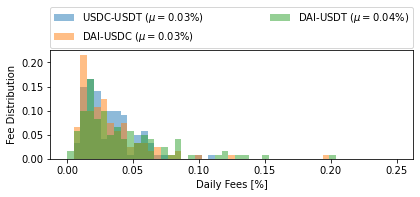}
    \caption{stable pairs} \label{fig:feestable}
  \end{subfigure}%
    
  \begin{subfigure}[b]{\linewidth}
    \includegraphics[scale=0.53,right]{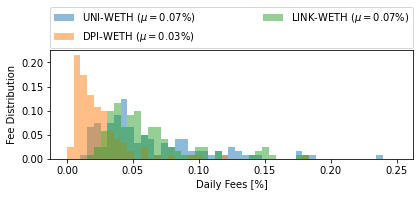}
    \caption{normal pairs} \label{fig:feenormal}
  \end{subfigure}%
  
  \begin{subfigure}[b]{\linewidth}
    \includegraphics[scale=0.53,right]{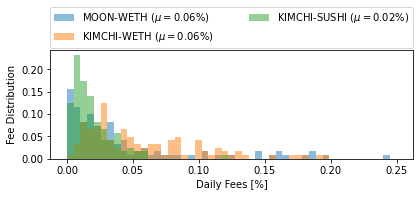}
    \caption{exotic pairs} \label{fig:feeexotic}
  \end{subfigure}

\caption{Daily percentage fees of sample pairs in three categories over a four month period. The histograms are normalized, i.e., the  sum of the bar heights of each data set are equal to one.} \label{fig:fee}
\end{figure}

\autoref{fig:fee} illustrates the fee distribution observed in the nine sample pools. While we can detect differences in the distributions of the fees collected in the various pools, there is no apparent pattern between the different types of pools. The fees collected by liquidity providers appear to depend more on the specific pair than on whether the pair is stable, normal, or exotic. 

\begin{figure}
  \begin{subfigure}[b]{\linewidth}
    \includegraphics[scale=0.53,right]{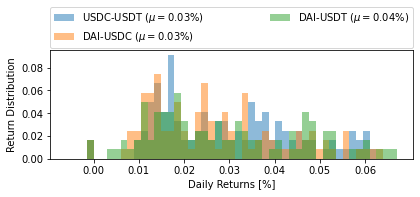}
    \caption{stable pairs} \label{fig:returnstable}
  \end{subfigure}%
    
  \begin{subfigure}[b]{\linewidth}
    \includegraphics[scale=0.53,right]{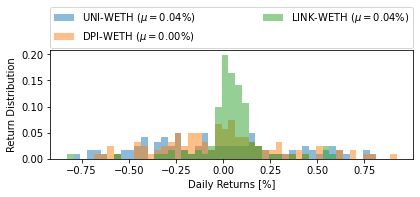}
    \caption{normal pairs} \label{fig:returnnormal}
  \end{subfigure}%
  
  \begin{subfigure}[b]{\linewidth}
    \includegraphics[scale=0.53,right]{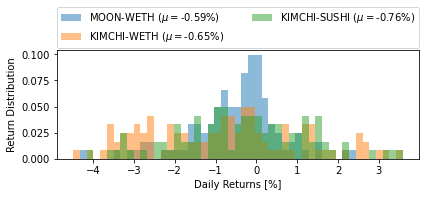}
    \caption{exotic pairs} \label{fig:returnexotic}
  \end{subfigure}

\caption{Daily percentage returns of sample pairs in three categories over a four month period. The histograms are normalized, i.e., the sum of the bar heights of each data set are equal to one.} \label{fig:return}
\end{figure}

Turning to the return rate observed in our sample pools, \autoref{fig:return}, we clearly observe patterns between the three different types of pools. Due to the previously observed negligible influence of impermanent loss on the profits expected in stable pairs, the daily returns in stable pools are rarely, if ever, negative -- \autoref{fig:returnstable}. Thus, all having positive daily average returns of around 0.03\%. For the normal pairs, \autoref{fig:returnnormal}, we observe significantly higher volatility in the returns, accompanied by both higher (0.04\% for $LINK\rightleftharpoons WETH$) and lower (0.00\% for $DPI\rightleftharpoons WETH$) daily returns than previously seen for the stable pairs. Finally, the exotic pairs seen in \autoref{fig:returnexotic} are characterized by even larger return volatility and very negative daily returns (-0.76\% for $KIMCHI\rightleftharpoons SUSHI$).

\begin{figure}
\centering
    \includegraphics[scale=0.51]{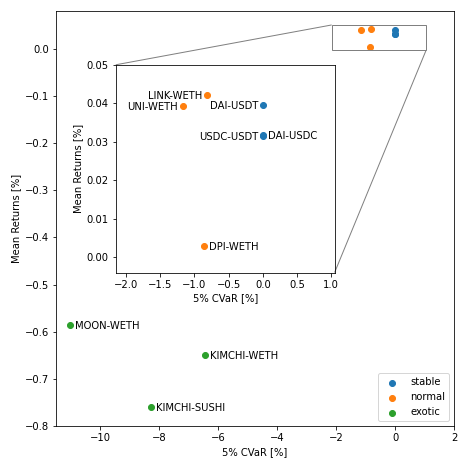}
    \caption{Risk and return comparison. We plot mean daily returns and the daily 5\% CVaR for nine sample pools. Providing liquidity for exotic tokens is beyond just risky.}
    \label{fig:risk}
\end{figure}

To summarize the risks and returns associated with providing liquidity, we turn to \autoref{fig:risk}. There we plot the daily mean returns observed in each sample pool against the daily \textit{conditional value at risk} (CVaR). CVaR is a coherent risk measure that represents the average worst-case scenario. More precisely, CVaR at 5\% level is the expected return on an investment in the worst 5\% of cases~\cite{rockafellar2000optimization}. We use historical data to calculate the daily mean and CVaR of the returns. Analyzing \autoref{fig:risk} allows us to extract patterns emerging between the three different kinds of pools that in-line with our previous observations. In general, the three exotic pairs have the highest risk, measured through the CVaR,  as well as the lowest average return associated with them. Exotic pairs make an inefficient investment to liquidity providers, as all three exotic pairs are dominated, i.e., higher return and lower risk, by each of the remaining pairs. For the normal pairs, on the other hand, the picture is more complex. While the $DPI\rightleftharpoons WETH$ pair is dominated by all stable pairs, the remaining two normal pairs are not. $LINK\rightleftharpoons WETH$ and $UNI\rightleftharpoons WETH$ exhibit higher returns than at least the majority of stable pairs, but with that also carry higher risks. All three stable pairs are very similar in terms of risk and return, with $DAI\rightleftharpoons USDT$ having a slight edge over the others in return. Stable and normal pairs may provide attractive opportunities for liquidity providers depending on their individual risk tolerance and return expectations.


\section{Movement Between Liquidity Pools}

As profits and risks associated with providing liquidity vary widely across different pools, we analyze how users redistribute their investments across different pools, i.e., how they move their assets from one pool to another. We investigate the movement of liquidity providers with the goal of better understanding their motivations.

\begin{figure}
  \begin{subfigure}{\linewidth}
    \includegraphics[scale=0.53,right]{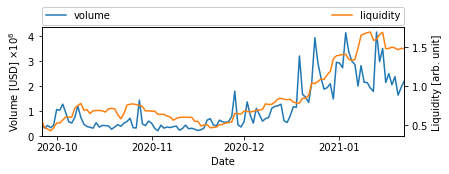}
    \caption{$USDC\rightleftharpoons USDT$} \label{fig:stablecorr}
  \end{subfigure}%
    
  \begin{subfigure}{\linewidth}
    \includegraphics[scale=0.53,right]{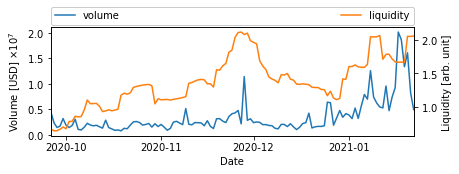}\vspace{-0.2cm}
    \caption{$LINK\rightleftharpoons WETH$} \label{fig:normalcorr}
  \end{subfigure}%
  
  \begin{subfigure}{\linewidth}
    \includegraphics[scale=0.53,right]{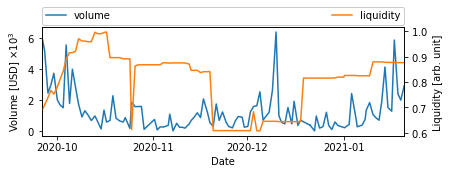}\vspace{-0.2cm}
    \caption{$KIMCHI\rightleftharpoons SUSHI$} \label{fig:exoticorr}
  \end{subfigure}

\caption{Volume and liquidity in three sample pools.} \label{fig:vollp}
\end{figure}

In \autoref{tab:corr} we record the correlation between the pool liquidity and other pool characteristics, namely volume, and token prices in USD. The correlations are calculated for daily data. For all stable pairs, \autoref{tab:corrstable}, volume correlates highly with pool liquidity. This strong correlation is expected, as liquidity provider's returns in stable pools are directly related to the fees accumulated (\autoref{fig:evstable}), which are in turn proportional to the volume. The high correlation between liquidity and trading volume indicates the dynamic balance of the liquidity providers in the market. When the trading volume increases, liquidity providers will earn more fees, then the high return attracts more liquidity into the pools. Symmetrically, when the trading volume decreases, liquidity will withdraw their funds from the pools, allowing the remaining liquidity providers to earn sufficient benefits again.
Thus, the correlation between liquidity and trading volume appears natural. We see that liquidity tracks volume in stable pools in \autoref{fig:stablecorr}, where the daily volume and liquidity are visualized for $USDC\rightleftharpoons USDT$. 

The data shown in \autoref{tab:corrnormal} for normal pairs and  \autoref{tab:correxotic} for exotic pairs reveals a less obvious picture. The liquidity in all but one of the normal and exotic pairs appears largely uncorrelated with the volume -- $LINK\rightleftharpoons WETH$ being the exception. \autoref{fig:normalcorr} shows that while liquidity and volume are somewhat correlated for the $LINK\rightleftharpoons WETH$ pair, the link between them is not as apparent as for the stable pairs. In the remaining pools, an example is shown with $KIMCHI\rightleftharpoons SUSHI$ in \autoref{fig:exoticorr}, liquidity and volume are uncorrelated. The lack of correlation might partially be due to generally low volume and low liquidity but could also stem from the less predictable returns in normal and exotic pairs. Thus, liquidity providers might pay less attention to the volume when adding or removing liquidity. In general, the price of the tokens appears uncorrelated with the liquidity. While we do not necessarily expect a strong correlation, the token price influences the liquidity providers' returns via the impermanent loss. Thus, liquidity providers do not seem to react to price changes, indicating that either they hope for the ratio to recover to previous value or they are insensitive to the effects of the impermanent loss.  
\begin{table}
    \vspace{0.2cm}
    \centering
    \scriptsize 
    \begin{subtable}{\linewidth}
    \centering
    
    \begin{tabular}{@{}p{1.6cm}  p{1.9cm}  p{1.9cm}  p{1.9cm}  @{}}
    	    \toprule
    		\textbf{} & \textbf{\tiny{USDC$\rightleftharpoons$ USDT}} &\textbf{\tiny{DAI$\rightleftharpoons$ USDT}} &\textbf{\tiny{DAI$\rightleftharpoons$ USDC}}\\
    		\midrule
            \textbf{Volume} & 0.82 &0.80&0.82\\
            \textbf{Price token 0} & 0.09 & 0.03&0.09\\
            \textbf{Price token 1} & 0.09 &  0.07& 0.13\\
            \bottomrule
           
    	\end{tabular}
    	 
    \caption{stable pairs} \label{tab:corrstable}
    \end{subtable}
    \\
    \begin{subtable}{\linewidth}
    \centering
    
    \begin{tabular}{@{}p{1.6cm}  p{1.9cm}  p{1.9cm}  p{1.9cm} @{}}
	    \toprule
		\textbf{}  &\textbf{\tiny{UNI$\rightleftharpoons$ WETH}} &\textbf{\tiny{DPI$\rightleftharpoons$ WETH}}& \textbf{\tiny{LINK$\rightleftharpoons$ WETH}}\\
		\midrule
        \textbf{Volume}  & 0.00 &0.14&0.48\\
        \textbf{Price token 0} &-0.31  & -0.29 &0.71\\
        \textbf{Price token 1} &-0.36 &-0.16 & 0.67 \\
       
        \bottomrule
        
	\end{tabular}
    
    \caption{normal pairs} \label{tab:corrnormal}
    \end{subtable}
    \\
    \begin{subtable}{\linewidth}
    \centering
    
    \begin{tabular}{@{}p{1.6cm}   p{1.9cm} p{1.9cm}  p{1.9cm}  @{}}
	    \toprule
		\textbf{} & \textbf{\tiny{MOON$\rightleftharpoons$ WETH}} &\textbf{\tiny{KIMCHI$\rightleftharpoons$ WETH}}&\textbf{\tiny{KIMCHI$\rightleftharpoons$ SUSHI}}\\
		\midrule
        \textbf{Volume} & 0.01& -0.20&-0.05\\
        \textbf{Price token 0} &0.31 &-0.50 &0.08 \\
        \textbf{Price token 1} & -0.11&  0.51 &  -0.05\\
        \bottomrule
        
	\end{tabular}
    
    \caption{exotic pairs} \label{tab:correxotic}
    \end{subtable}
    \caption{Correlation between the liquidity in the pool and various pool characteristics. The correlations are recorded for daily data.}
    \label{tab:corr}
\end{table}

We further look at the movement of liquidity between pools. When observing liquidity movements, we consider the entire data set. We record a movement if the same address removes liquidity from one pool and adds liquidity to another pool within 6000 blocks -- roughly a day. Further, we restrict our analysis to the pools with the most liquidity movements (mint and burn events) in the following and only consider pools with more than 5000 liquidity movements. 

\begin{figure}
\centering
\includegraphics[scale=0.51]{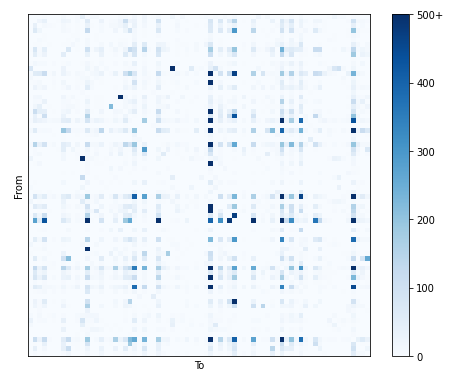}\vspace{-0.2cm}
    \caption{The number of directional movements between the 72 most active pools. The pools are ordered smallest to largest by their average size, i.e., liquidity in the pool, since the day of their creation. We cap the number of movements at 500 for better visibility.}
    \label{fig:movement}
\end{figure}

In \autoref{fig:movement} we plot a colormap of the movement between the 72 pools most active pools. We order the pools smallest to largest by their average size, i.e., liquidity in the pool, since their creation. For better visibility, all values are capped at 500. We immediately draw three conclusions from \autoref{fig:movement}. First, the movements of liquidity in Uniswap are rare. Additionally, the matrix appears symmetric. In other words, if providers move liquidity from one pool to another, there also seem to be providers that move liquidity in the opposite direction at similar numbers. Lastly, the number of movements don't appear to correlate with the pool size. The ordered pool pair with the most movements is $DAI\rightleftharpoons WETH \rightarrow WETH\rightleftharpoons SURF$ with a total of 13389 movements. We count less than 3500 movements for all remaining ordered pairs.

\begin{figure}
  \begin{subfigure}{\linewidth}
    \includegraphics[scale=0.52]{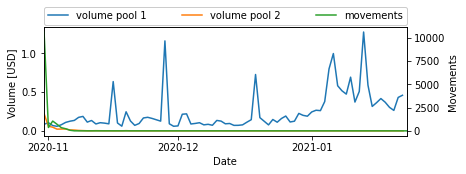}\vspace{-0.2cm}
    \caption{$DAI\rightleftharpoons WETH \rightarrow WETH\rightleftharpoons SURF$} \label{fig:move1}
  \end{subfigure}%
  
  \begin{subfigure}{\linewidth}
    \includegraphics[scale=0.52]{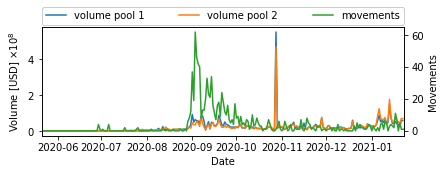}\vspace{-0.2cm}
    \caption{$USDT\rightleftharpoons WETH  \rightarrow  USDC\rightleftharpoons WETH$} \label{fig:move2}
  \end{subfigure}
  
  \begin{subfigure}{\linewidth}
    \includegraphics[scale=0.52]{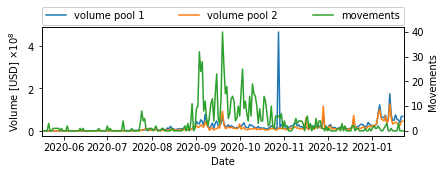}\vspace{-0.2cm}
    \caption{$USDC\rightleftharpoons WETH \rightarrow  DAI\rightleftharpoons WETH$} \label{fig:move3}
  \end{subfigure}
\caption{Directional liquidity movements and volume for a set of ordered pool pairs.} \label{fig:move}
\end{figure}

When looking at the data of some ordered pool pairs with many movements in more detail, we observe the emergence of two patterns in \autoref{fig:move}. While movements between $DAI\rightleftharpoons WETH \rightarrow WETH\rightleftharpoons SURF$ all occur within a rather short period and appear to be driven by an individual event, the movements we observe between $USDT\rightleftharpoons WETH  \rightarrow  USDC\rightleftharpoons WETH$ and $USDC\rightleftharpoons WETH \rightarrow  DAI\rightleftharpoons WETH$ happen over a longer period. However, even though we observe different patterns, movements appear to relate mostly to liquidity mining. The single spike in between $DAI\rightleftharpoons WETH \rightarrow WETH\rightleftharpoons SURF$ (\autoref{fig:move1}) coincides with the time at which liquidity mining for $SURF$ started in the $WETH\rightleftharpoons SURF$ pair\footnote{https://www.reddit.com/r/CryptoMoonShots/comments/jivadf/surf\_finance/}. Further, the movement between the $USDT\rightleftharpoons WETH  \rightarrow  USDC\rightleftharpoons WETH$ and $USDC\rightleftharpoons WETH \rightarrow  DAI\rightleftharpoons WETH$ reaches its peaks around the UNI liquidity mining period, 18th September 2020 to 17th November 2020\footnote{https://uniswap.org/blog/uni/}. In conclusion, the movement of liquidity does not appear too common among liquidity providers unless driven by external motivations, such as liquidity mining.

\section{Conclusion}

This paper addresses three fundamental problems in understanding liquidity providers in DEXes: Who provides liquidity in DEXes? What are the returns and risks of providing liquidity? How do liquidity providers react to market changes? 

Our analysis suggests that users of cryptocurrency ecosystems have gradually become interested in providing liquidity in DEXes. However, users still act with caution -- preferring to participate in few liquidity pools. We find that the returns and losses of providing liquidity in different types of pools vary a lot. Stable pools enable a seemingly risk-free and profitable investment opportunity compared to constant-mix portfolios. Providing liquidity in exotic pools, on the other hand, appears to perform much worse than the corresponding constant-mix portfolios. Reacting to the unique return opportunities and risks, liquidity providers perform different trading strategies across pool categories: they respond to trading volume changes in stable pools and pay less attention to market changes of normal and exotic pools. Besides market indicators, liquidity providers are also motivated by external market factors, i.e., liquidity mining activities, to redistribute their liquidity investments.

In this paper, we extend the research scope of liquidity provider's behavior in DEXes from theoretical analysis to empirical studies. By studying the behaviors of liquidity providers on the Uniswap market, our work provides a comprehensive insight into the new trading options in the cryptocurrency ecosystems for users. More, it inspires future work aimed to better understand  DEXes market mechanisms.

\bibliographystyle{ACM-Reference-Format}
\bibliography{reference}

\end{document}